\documentclass[sigconf]{acmart}

\AtBeginDocument{%
  \providecommand\BibTeX{{%
    \normalfont B\kern-0.5em{\scshape i\kern-0.25em b}\kern-0.8em\TeX}}}

\copyrightyear{2022}
\acmYear{2022}
\setcopyright{acmlicensed}\acmConference[AIES'22]{Proceedings of the 2022 AAAI/ACM Conference on AI, Ethics, and Society}{August 1--3, 2022}{Oxford, United Kingdom}
\acmBooktitle{Proceedings of the 2022 AAAI/ACM Conference on AI, Ethics, and Society (AIES'22), August 1--3, 2022, Oxford, United Kingdom}
\acmPrice{15.00}
\acmDOI{10.1145/3514094.3534181}
\acmISBN{978-1-4503-9247-1/22/08}

\begin{document}

\title{Outsider Oversight: \\ Designing a Third Party Audit Ecosystem for AI Governance}

\author{Inioluwa Deborah Raji}
\email{rajiinio@berkeley.edu}
\affiliation{%
  \institution{University of California, Berkeley}
  \city{Berkeley}
  \state{CA}
  \country{USA}
}

\author{Peggy Xu}
\affiliation{%
  \institution{Stanford University}
  \city{Stanford}
  \state{CA}
  \country{USA}}
\email{peggyxu@stanford.edu}

\author{Colleen Honigsberg}
\affiliation{%
  \institution{Stanford University}
  \city{Stanford}
  \state{CA}
  \country{USA}}
\email{colleenh@law.stanford.edu}

\author{Daniel Ho}
\affiliation{%
  \institution{Stanford University}
  \city{Stanford}
  \state{CA}
  \country{USA}}
\email{dho@law.stanford.edu}


\begin{abstract}
Much attention has focused on algorithmic audits and impact assessments to hold developers and users of algorithmic systems accountable. But existing algorithmic accountability policy approaches have neglected the lessons from non-algorithmic domains: notably, the importance of interventions that allow for the effective participation of third parties.
Our paper synthesizes lessons from other fields on how to craft effective systems of external oversight for algorithmic deployments. First, we discuss the challenges of third party oversight in the current AI landscape. Second, we survey audit systems across domains – e.g., financial, environmental, and health regulation – and show that the institutional design of such audits are far from monolithic. 
Finally, we survey the evidence base around these design components and spell out the implications for algorithmic auditing. We conclude that the turn toward audits alone is unlikely to achieve actual algorithmic accountability, and sustained focus on institutional design will be required for meaningful third party involvement.
\end{abstract}







\maketitle

\section{Introduction}
Some of the most stunning findings of algorithmic bias and harm have been uncovered through the exceptional efforts of third parties. Regulators, academics, civil society, investigative journalists, specialized law firms and more have already begun to form a network of external actors participating in the  informal and formal third party oversight of algorithmic systems. Together, these watchdogs continue to participate in consequential studies to analyze the biased outcomes of models used in domains as varied as content curation~\cite{sweeney2013discrimination}, online advertising~\cite{kofman2019facebook}, tenant screening~\cite{rosen2021racial,markup1kirchner2020access}, child welfare~\cite{chouldechova2018case}, criminal justice~\cite{eckhouse2019layers,angwin2016machine,buolamwini2018gender,snow2018,grother2019face}, hiring~\cite{raghavan2020mitigating}, immigration~\cite{FOX2}, education~\cite{FOX1} or public health~\cite{coston2021leveraging, obermeyer2019dissecting}. 
The impact of this work is undeniable --- leading to voluntary moratoriums~\cite{heilweil2020big}, and several successful lawsuits~\cite{sherwinfacebook, spinks2019contemporary}; nation-wide waves of regulatory action and public outcry~\cite{fban}; congressional hearings~\cite{markupimpact}; and the recalls or radical redesigns of the deployed algorithms~\cite{wilson2021building, raji2019actionable, raji2020saving, chouldechova2018case}.

However, despite its clear impact, less attention has been paid to the crucial role of third-party oversight in policy proposals that mention algorithmic auditing more generally. Developments such as the General Data Protection Regulation (GDPR)~\cite{voigt2017eu}, the proposed Algorithmic Accountability Act~\cite{maccarthy2019examination}, and several of the state-wide and municipal proposals throughout the U.S.~\cite{maurernew2021} tend to focus instead on \emph{internal audits} commissioned, executed, paid for and controlled by the companies being targeted.  

Many experts nevertheless applaud such developments.  Kearns and Roth note that “more systematic, ongoing, and legal ways of auditing algorithms are needed”~\cite{kearns2020ethical}. Kai-Fu Lee even asserted that such internal audits may be “better than regulations,” and pointed to the exemplar of Environmental, Social, and Governance (ESG) reporting and auditing for corporate social responsibility (CSR).\footnote{Eye on AI Podcast, Aug. 11, 2021, at 15:00 (“It’s impossible for a government to basically evaluate all companies in person so \dots audits would make sense \dots ESG has generally been successful \dots in guiding positive behavior in corporations \dots  If there are ESG third party watch dogs \dots and if shareholders started to care  \dots that creates a self-monitoring mechanism that could work perhaps even better than regulations.”)} 

While the sharp rise in ESG investment strategies is indeed impressive, ESG in many ways challenges, not supports, the turn toward the audits of algorithmic and artificial intelligence (AI) branded products. Much like “AI ethics,” the notion of ESG is expansive – encompassing global warming, human rights, labor practices, executive compensation, diversity, equity, and inclusion and more. The potential costs of diffuse ESG audits are high. 
And due to the ambiguity of what is being audited, ESG certifications risk becoming “cheap talk,” rubber stamping practices without in fact promoting social responsibility.\footnote{\citet[][noting “many companies seem to be doing the minimum necessary to say they obtained assurance on some of their sustainability reporting and many assurance providers are using standards that are so general and discretionary as to provide the consumers of ESG reporting little practical assurance.”]{rossseymour}} A large literature has pointed to these challenges in ESG investing~\cite{rossseymour,o2021esg,eccles2012need, kim2020analyzing,boffo2020esg,pollman2019corporate,kotsantonis2019four}.  
There is a clear precedent for situations of auditors compromised by selection, payment, or employment by auditees; auditors struggling to secure access; auditors clumsily navigating diffuse performance standards; and auditees that may elect to share or conceal findings otherwise hidden behind a nondisclosure agreement.

The naive move toward AI audits – without considering facets like third-party auditor scope, independence, access, standards, and transparency – risks ignoring these lessons.
Like in ESG, conflicts of interest can compromise the quality of third-party audits of algorithmic systems. 
Auditing the automated decision systems of a large technology platform can rival the complexity of auditing for human rights violations in a large supply chain.
The standards and scope of an algorithmic audit may be unclear - leading to assessments that are unfocused and expensive. The mediation of auditor access is difficult, and the identification of audit targets and scope uncertain. 
Furthermore, audited companies could outright ignore or overlook audit results. 
Worse yet - as was observed in the AI hiring tool industry~\cite{sloaneaudit,sloane2021silicon} - even well-executed algorithmic audit results may get co-opted by audit targets and used to disguise deeper failures. 
The controversial facial recognition company Clearview AI, for example, claimed to be “100\% accurate across all demographic groups according to the ACLU’s facial recognition accuracy methodology,” but failed to address the ethical and legal challenges that accompany its sourcing of face data, which has been subject to numerous legal actions (e.g., in California, Illinois, the European Union)~\cite{ACLUClearview}. 

In sum, not all audits are created equally. Ultimately, much more is required to make algorithmic audits work, including the intentional design of an audit ecosystem enabling the effective participation of third parties. Existing evidence from other regulatory fields suggests that effective target identification mechanisms, independence in selection and compensation, access to data and AI systems, clear and well-scoped audit standards, and post-audit report transparency will be critical to ensure that audits promote, rather than degrade, overall accountability. This article articulates the current challenges for third party oversight (Section 2), surveys the institutional design of audit systems in non-algorithmic domains (Section 3), and synthesizes the evidence base to draw implications for how to design an effective, robust, and scalable AI audit ecosystem (Section 4).  We argue that much of what is missing in the policy discussion of algorithmic auditing — and especially third-party algorithmic auditing — are lessons that can be drawn from existing empirical literature on non-algorithmic audits. Due to practical limitations, we confine our analysis mainly to the U.S. context.

\section{The Current U.S. AI Audit Policy Landscape}

We proceed by clarifying some terminology and then assess the current landscape of AI audits, with a focus on the U.S. context. In this article, we use ``algorithm", and related terms such as ``automated decision-making systems'' interchangeably with ``artificial intelligence (AI)'' branded products.

\subsection{Terminology}

Audits evaluate performance relative to expected behavior, as part of a broader accountability process~\cite{power1997audit}. Audits can be qualitative or quantitative, as system expectations can be articulated and assessed in either mode. The scope of an audit does not always allow for a broader view on the downstream impacts of a system or more open ended reflections on the consequences of deployment beyond the direct outcomes of the system itself. Instead, the audit is most focused on the comparison of the reality of the product's behavior with respect to clearly articulated expectations, standards and claims.

By convention, audits are classified into three types~\cite{cochran2015iso}. “First party” 
audits are conducted by a company of its own products. Many AI ethics teams can be conceived of as fulfilling a first-party audit role. “Second party” audits are performed by a contractual counterparty or an entity hired by that contractual counterparty. Second party audits typically ensure compliance with contract terms.
As applied to AI, public sector procurement of AI products may require a second party audit to assess bias, as has been the case for legislative proposals to regulate facial recognition technology. “Third party” audits are conducted by ostensibly independent parties engaged specifically to conduct the audit, typically subject to pre-determined auditing standards. Facebook, for instance, engaged the non-profit organization Business for Social Responsibility to conduct a human rights impact assessment of Facebook’s entrance into Myanmar in 2018~\cite{latonerohuman2021}. Another example would be regulators at the U.S. Food and Drug Administration auditing a medical AI tool prior to market entry. We note that while third party oversight by non-profit organizations, journalists, and academics have, as noted above, played critical roles in exposing weaknesses to AI systems, such forms of oversight are not conventionally considered audits per se. 

Another distinction is between internal and external oversight. Internal oversight is conducted by  stakeholders with a direct employment or contractual relationship with the auditee~\cite{raji2020closing}.  External oversight is conducted by third parties with no direct employment or contractual ties to the audit target~\cite{raji2019actionable}. Such external parties, which may be public (ie. tied to governmental or other public-interest institutions) or private sector entities, may have no direct access to the audit target and design the audit with different objectives beyond compliance.

We note that the terminology can present uncertainties.
Some might label a consultant or contractor hired by a company to conduct an audit to be a ``third party'' audit, but such an audit may not be considered a form of ``external'' oversight due to the compensation scheme. Others may use the term ``external" and ``third party'' interchangeably. These uncertainties are precisely the reason to focus on the specific institutional arrangements of an audit scheme, as we articulate below. 

\vspace{-3mm}
\subsection{The Outsized Focus on Internal Audits}

With these primitives in mind, it is worth noting several major limitations of the current AI audit landscape. 

First, many AI policy developments tend to focus on 
articulating some guidance on what an internal audit should include and potentially enforcing these as expectations for a bare minimum “form” an audit is meant to take. 
These instructions around audit execution are ultimately attempts at control on audit quality and meaningfulness for accountability, but also reveal the assumption underpinning many of these proposals that \emph{internal} parties are expected to be the ones conducting compliance audits. 
For instance, the New York City Council mandate for hiring tools specifies the details of a “bias” audit~\cite{maurernew2021}. The Algorithmic Accountability Act of 2022 lists out specific minimum requirements for what is to be included in the required algorithmic impact assessment (AIAs), including a breakdown of evaluation and documentation expectations. Similarly, the Information Commissioner's Office released a report of “auditability” guidelines for the implementation of impact assessments required as part of the enforcement of the General Data Protection Regulation (GDPR), describing in detail the types of graphs and calculations require to assess notions of accuracy, fairness and explainability~\cite{noauthorai2020}. All of these format and content oriented developments work to restrict and define what is to be included in the audit, focusing on concerns that are preferable for internal auditors to operationalize, and that requires a certain level of access or internal co-operation to achieve. 

Specifically, the assumption behind calls for Algorithmic Impact Assessments (AIAs) remains that audit activities are executed by internal stakeholders.  Most mainstream interpretations of the AIA continue to center its execution as the responsibility of the assessed organization~\cite{reisman2018algorithmic,selbst2021institutional}. Even in cases when a participatory approach is considered, the responsibility of consulting with external perspectives when considering impacts to articulate in an AIA is still framed as the responsibility of internal stakeholders to coordinate and interpret these interactions~\cite{metcalf2021algorithmic}. Thus, AIAs are often practically executed by similar stakeholders as internal audits, with broader regulatory implications and a purpose framed towards prompting and gauging institutional reflexivity and due diligence, in addition to the audit goals of performance assessment with respect to articulated expectations~\cite{lovelace2020examining}. Unlike most audits, AIAs are not as strictly defined and tend to include more open-ended inquiry into the impacts of a particular algorithmic deployment~\cite{moss2021assembling}. The scope of an AIA is thus potentially broader than a specific product feature, and the assessment goes beyond just evaluation measurements, to include more reflective articulations of the trade-offs and decision-making inherent in the release and development of a particular algorithmic system. 


Second, the emphasis on internal audits can omit concerns by communities impacted by algorithmic deployments. Internal accountability approaches are developed in reference to a limited set of compliance requirements and self-determined AI principles~\cite{raji2020closing}. This tends to prioritize at most the concerns of regulators or customers, primarily to reduce liability risk, which may not align with core concerns of impacted populations. While helpful in aligning internal stakeholders, such an approach is not likely to identify or address the main concerns of external stakeholders~\cite{mittelstadt2019principles, raji2020closing}, particularly if that affected population or its representatives are systematically neglected or under-considered as affected non-users. For instance, companies developing algorithmic policing tools naturally prioritize the needs of the police users of their products, even though the community whose safety is impacted by the use of such tools may have different expectations and preferences~\cite{mcgrory2020targeted}. In the same vein, regulatory and institutional interpretations of hate speech often differs dramatically from the understanding of those experiencing hate online~\cite{kumar2021designing}, leading to deployed moderation tools that further marginalize and censor impacted users rather than protect them~\cite{diaz2021double}. Internal auditors can also sometimes direct their attention to focus audits on concerns that are easier for them to measure and address --- for example, Twitter~\cite{twitter1}, Facebook~\cite{hao2021facebook} and Hirevue~\cite{hirevueAP} have all been critiqued for promoting a myopic focus on fairness issues while simultaneously diverting attention from other concerns related to the more difficult to formalize challenges of privacy, misinformation and product validity~\cite{engler2021independent}.   

Thirdly, there is acute potential for conflicts of interest with first or second party audits. For instance, Google researchers on the internal AI team were dismissed and blocked from publishing critiques on the large-scale language models~\cite{hao2020wetimnit1, ebell2021towardstimnit2}, which would later be revealed to be critical to the company's future product roadmap~\cite{fedus2021switch, googl1, googl2}.  Similarly, the warnings from Facebook researchers on addressing mounting bias issues and misinformation campaigns was reported to have been internally suppressed~\cite{oremus}.  Even the work of hired consultants can be manipulated when the publication process is mediated by the company itself, as demonstrated by the conflicted reporting on the Hirevue audit by consulting company ORCAA~\cite{engler2021independent}. As a result of this conflict of interest, the ability of internal stakeholders or hired parties to properly scrutinize, honestly describe and effectively alter the algorithmic development process or its outcomes is limited when the incentives to do so directly contradict corporate preferences.

Fourth, even when external auditing is contemplated, the focus is on a narrow set of potential auditors. For instance, the Digital Services Act restricts external access privileges to academic researchers. Similarly, the Algorithmic Accountability Act in 2022, and the Algorithmic Justice and Online Platform Transparency Act of 2021~\cite{noauthorsenator2021} contemplate obligations on regulated parties with enforcement by the Federal Trade Commission (FTC). Other proposals put the onus on regulatory agencies to serve the primary role of external auditor~\cite{kearns2020ethical}, but regulatory capacity to conduct such audits is currently lacking~\cite{engstrom2020algorithmic}. In recognition of this capacity gap, the new Algorithmic Accountability Act of 2022 calls for 75 new staff positions for the Federal Trade Commission. 

Last, the limited audit capacity has led to many instances of audit failures. Before 2019 and the intervention of the Algorithmic Justice League’s Gender Shades audit~\cite{buolamwini2018gender}, the U.S. National Institute of Technology (NIST) failed to measure demographic differences in their regular Facial Recognition Vendor’s Test (FRVT), thus overlooking for many years that “Asian and African American people were up to 100 times more likely to be misidentified than white men”~\cite{grother2019face}. Similarly, the U.S. Food and Drug Administration has faced challenges in assessing AI-enabled medical devices, largely deferring to company benchmarks on limited, retrospective data~\cite{wu2021medical}. 

In sum, while policy attention has been turning toward audits, serious limitations characterize the current AI audit landscape. A future where AI audits are conducted by company employees and consulting companies alone fails to learn the lessons of how third parties have generated robust oversight of AI systems. Such systems are diverse, impacting a range of industries and stakeholders, and it will require more third party participants, representing a variety of communities and making use of a wide set of tools, to concretely assess a wider range of the harms involved for the implicated stakeholders.

\vspace{-3mm}
\section{The Institutional Design of Audit Systems}

Given the limitations of the current AI audit landscape, we now proceed to examine what we can learn from audit systems in other domains. Our goal is to examine other industry successes and failures to consider how to best approach the design of an effective audit ecosystem. 
We first articulate main differences in how audit systems are designed. We then survey a range of audit systems and illustrate how such differences are manifested. 

We break down the institutional design features into five main considerations: (i) \textbf{Target Identification \& Audit Scope:} The grounds through which the audit subject is identified, and the audit criteria are scoped; (ii) \textbf{Auditor Independence:} The degree to which auditors are able to operate without the influence of potentially conflicted stakeholders (e.g., auditor selection, compensation, and rotation requirements); (iii) \textbf{Auditor Privileges:} Any special permissions or protections provided to auditors to facilitate the audit process, and protect auditors against retaliation (e.g., access and legal immunity); (iv) \textbf{Auditor Professionalization \& Conduct Standards:} Articulated rules for audit methodology and auditor qualifications (e.g., accreditation, certification); (v) \textbf{Post-audit actions:} Requirements on the auditor, audit target, or other stakeholders to act on audit results (e.g., disclosure). 

We now apply this typology to a range of existing audit schemes. We note that the domains are diverse, ranging from financial services, environmental protection, telecommunications, transportation, and healthcare. Some of these audit schemes already intersect with AI products -- the National Transportation Safety Board (NTSB) has provided third party analysis of self-driving car crashes; the Food and Drug Administration (FDA) is already approving AI-enabled medical devices; and many current internal audit templates derive from documentation requirements in these other industries. Some AI products, in that sense, are already subject to incumbent audit policies. 

We summarize our survey of audit systems in Table \ref{ecosystem}. Rows present distinct audit systems (for descriptions, see Appendix Table \ref{audit-descriptions}) and columns present the five main design features (for details see Appendix Table \ref{ecosystem-features}). We group audit systems by whether they are first, second, or third party audits.  

We highlight several patterns. First, we note that nearly all audit programs have specific standards and rules that help to scope the audit. Audits to determine whether a good producer meets organic labeling, for instance, are conducted subject to a federal standard~\cite{agri_report}.
Second, we note that significant variation exists within the third party audit category, which may be conducted by public agencies (e.g., nursing home inspections) or private parties (e.g., children’s toys), which typically also determines whether auditors are compensated by the audited party.  Cases in the third party panel are roughly sorted from schemes that are closest to first and second party audits to schemes that are closest to full-blown public inspections. Due to limited regulatory capacity, public inspections appear more likely to be initiated by some complaint or risk-based heuristic, rather than being universally mandated. Third, all audit systems have some mechanism of access to otherwise confidential information: e.g., access to enter a facility to inspect the premises and records or the reporting or access to financial statements. This stands in sharp contrast to third party oversight of AI products, where access has proven challenging. Fourth, accreditation schemes exist for all instances where the auditee selects and compensates the third party auditor. Rotation requirements, which require auditors to rotate across audit targets to ostensibly foster independence, are employed in a minority of cases. Last, in many cases, the audit scheme contemplates post-audit actions, such as recommendations for remediation and disclosure. 

What this survey highlights most of all is that there are many varieties of audits. A third party audit where (a) the auditor is selected and compensated by the auditee (b) to conduct an audit of a scope and access determined by the auditee, and (c) with limited post-audit actions may be functionally indistinguishable from first-party audit, and questionably effective in garnering the expected downstream accountability outcomes. 

\begin{figure*}
  \includegraphics[width=0.9\textwidth]{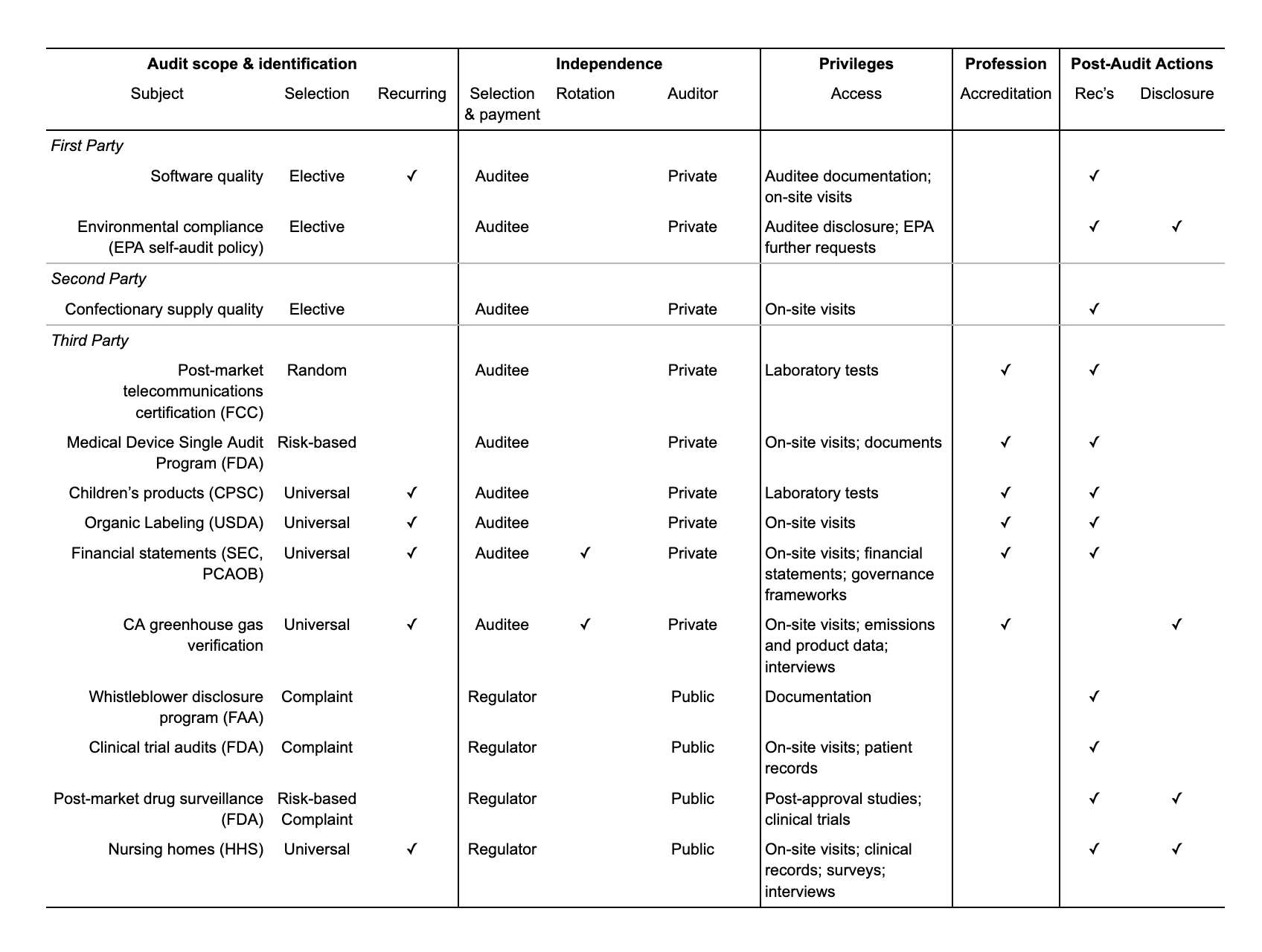} 
  \caption{This table classifies audit case studies along a range of different design features. “Subject” refers to what is being audited. “Selection” asks how the related harm was discovered and reported (e.g. whether the audit is mandated, voluntary, or complaint-initiated).  “Recurring” asks whether there is an ongoing or periodic auditing requirement. “Selection \& payment” refers to the entity that pays and selects the auditor. “Rotation” refers to whether auditors are required to rotate off of the auditing process after a certain period of time. “Access” explains how access is granted to auditors. “Accreditation” asks whether the auditing entity must be accredited, certified, or recognized by an official body before it is permitted to conduct the audit. “Rec’s” refers to whether the auditor has the power to make recommendations or changes to the audited entity. “Disclosure” asks whether the contents of the audit must be made available to the public in some form.}
  \label{ecosystem}
\end{figure*}


\section{The Evidence Base for Effective Audits}
We now examine the evidence base for which design choices may improve the overall effectiveness of an audit. We organize our discussion along the five design categories noted above. 
For each, we describe the main audit design considerations, present the evidence base from relevant social science and other industries, and develop implications for the AI context. 

\subsection{Target Identification \& Audit Scope}
Identifying audit targets and scope is critical to ensuring that audits are applied to relevant parties, and conducted for the right issues. 

\subsubsection{Evidence Base} 

\paragraph{\underline{Audit Selection}} Table \ref{ecosystem} shows that a fair number of audit schemes are universal (i.e., they are mandatory for all entities subject to the jurisdiction of the regulator). When audit resources are scarce, however, it may not be possible to select all parties for an audit.  In those instances, risk-based or complaint-based selection of audit targets can help to allocate attention where needed the most. Similar resource constraints prompted the development of “syndromic surveillance systems” in public health~\cite{henning2004syndromic}, adverse event reporting systems for pharmaceuticals~\cite{sakaeda2013data}, and financial fraud complaint databases~\cite{littwin2014process}, with the goal of providing a real-time signal about potential problems and thus focusing audit efforts. FDA’s “Adverse Event Reporting System,” for example, serves as a kind of post-marketing safety surveillance program for drugs and devices, flagging products with repeated complaints for secondary inspection. Other examples of incident reporting systems abound, including the FDA’s medical device incident database \cite{maude}, the Federal Aviation Administration’s Aviation Safety Reporting System~\cite{noauthoraviationnodate}, NIST’s National Vulnerabilities Database~\cite{noauthornationalnodate}, and the Consumer Financial Protection Bureau’s complaint system~\cite{plutaso2013}. These third-party reporting systems can help focus auditing resources on issues that matter most~\cite{tatonetti2012novel}.

Third-party reporting is most successful when third parties possess high quality information. When an individual’s income is reported to the IRS by third parties like banks and employers, for example, that individual is far more likely to be tax compliant than if they self-reported their income~ \cite{kleven2011unwilling,johnson2019federal}. Third-party reporting has a similar effect on small businesses~\cite{adhikari2021small}. Still, consumer complaints have proven to be a valuable signal when harms can be difficult to observe. One analysis of investigated nursing home complaints, for example, found that the complaints predicted performance at subsequent inspections~\cite{stevenson2006nursing}. Another research study found that the Vaccine Adverse Event Reporting System (VAERS) database is predictive of true safety concerns with newly released vaccines~\cite{geier2004review}. 

Complaint-initiated audits, however, also raise challenges. Voluntary reporting can be unreliable and create reporting bias~\cite{altenburger2019algorithms}. Residents in low-income and minority communities, for instance, are less likely to complain about street conditions or “nuisance” issues in a 311 system~\cite{kontokosta2021bias}. Similar biases have been found in recent consumer complaints to the Consumer Financial Protection Bureau (CFPB)~ \cite{porter2012complaint}—although early CFPB complaints, if anything, appeared to be higher from minority neighborhoods~\cite{ayres2013skeletons}.

Companies can themselves adopt incident reporting systems to raise user and public awareness of a system’s use in consequential scenarios. Similar to the “How Am I Driving?” sticker on vehicles for employee driver feedback \cite{strahilevitz2006s}, such prompts could nudge harm reporting for impacted stakeholders. These notices, however, can easily become overwhelming. Some have noted, for example, that the privacy disclosure pop-ups implemented in response to the EU's ePrivacy Directive have increased “consent fatigue” among users without meaningfully improving user privacy~\cite{sanchez2019can,trevisan20194}. In some cases, users were purposefully inundated with notices designed to induce them to consent by default~\cite{soe2020circumvention,nouwens2020dark,matte2020cookie,miyazaki2008online}. These types of informational disclosures can also distract regulators from more pressing problems. One study found, for instance, that restaurant compliance notices were not representative of actual food safety status, and pushed inspectors to spend outsize resources on adjudicating the accuracy of these notices rather than addressing safety hazards directly~\cite{ho2012fudging}. 

\paragraph{\underline{Audit Precision}} For an audit to successfully measure performance, its scope should be sufficiently delineated. Practitioners and researchers alike have noted the importance of making sure that the scope of an audit isn’t too broad or too narrow~\cite{karapetrovic2001audit}. If auditors are tasked with vague inspection mandates, their results may be difficult to understand or translate into enforcement actions or changed practices. Take the inconsistency of ESG climate change disclosures: Eccles et al. (2012) found that, despite the SEC’s interpretive guidance, there is significant variation between companies and within industries in climate change reporting, with some companies disclosing no information about climate change, others using boilerplate language in their disclosures, and still others including robust quantitative metrics like GHG emissions and energy use~\cite{eccles2012need}. As a result, the authors recommended that industries emphasize precision in determining audit jurisdiction and develop sector-by-sector standards for what metrics should be included in ESG climate reporting. The Financial Accounting Standards Board (FASB) uses materiality to “allow auditors to narrow the audit scope to focus on only the most important items." Some evidence suggests that auditing can help to standardize information for more consistent reporting~\cite{haapamaki2019research}.
 
\subsubsection{Implications for AI Auditing} 

\paragraph{\underline{Audit Selection}} The literature suggests that there may be benefits to establishing a \textbf{\textit{national incident reporting system}} to enable third parties to file complaints about algorithmic systems. Nonetheless, considerations for how to adjust for differences in reporting propensities and actively solicit minority perspectives for incident reporting will be a key component of the system's design. Much like how FDA’s system has enabled a wide range of techniques to surface issues, the incident reporting system can focus auditing resources on issues that matter most~\cite{tatonetti2012novel}. Regulators can use this information to identify specific priority targets and assess fees or penalties based on audit results. They could then incentivize third party auditors to inspect repeatedly flagged cases in this database through various means, such as RFPs or bug bounties.\footnote{This could resemble Twitter's algorithmic bias bug bounty \cite{twitter1}, except mediated by public actors, and not the audit target itself.}

The status quo is that the use of algorithms is rarely disclosed in a transparent manner, making it hard to identify audit targets in the first place~\cite{veale2018fairness}. As a result, it can be difficult for regulators and the public to identify incidents of algorithmic harm and the \textbf{\textit{public disclosure of algorithmic products in use}} is necessary to reveal such issues. Current strategies for harm discovery in AI involve ad hoc processes of notice and reporting, limited to situations of harm raised and escalated in public forums such as social media or the press. Third parties have already taken advantage of a range of interventions to identify and prioritize potential harms. At times, public procurement process disclosure requirements such as the Public Oversight of Surveillance Technology (POST) Act~\cite{richardson2019confronting} will require that certain (often public) institutions such as the New York Police Department (NYPD) publicly release information on which AI technologies are being used. System of Records Notices (SORN), which are intended to notify the public whenever a federal agency creates, modifies, or eliminates records involving personally identifiable information, may alert parties to potential uses of algorithms. Investigative journalism tools such as Algorithm Tips~\cite{trielli2017algorithm}, a database of algorithmic tools being used by public agencies, can also make information on algorithmic deployment more publicly accessible. However, even in instances when the algorithmic systems are themselves disclosed, it remains difficult to anticipate which of these deployments are linked to known incidents of harms, and thus challenging to prioritize cases or pursue targets proactively~\cite{raji2019actionable}. When communities identify issues with algorithmic deployments, it can be difficult for them to articulate or find concrete evidence for such harms~\cite{katell2019algorithmic}. Even when connected to representative organizations, such individuals or communities often lack the vocabulary and ability to flag the issues they observe~\cite{katell2020toward}. 

\paragraph{\underline{Audit Precision}} Audits should be \textbf{\textit{scoped precisely}}. The Digital Services Act is meant to regulate “recommender systems” developed by very large online platforms (VLOPs). The Algorithmic Accountability Act of 2022, meanwhile, applies to deployed “automated decision systems” (ADS) used in an “augmented critical decision process”. Although definitions are provided for these terms, there is little guidance for how regulators, auditors, and companies are meant to translate such descriptions into specific performance expectations. No auditor can audit an entire VLOP single-handedly, and the definition of ADS is so context-dependent and purposefully broad~\cite{richardson2021defining} that it does little to help scope an audit. One obvious place to start to narrow the scope is with existing law. Instead of asking, “Is there bias?”, an audit may more productively ask, “Does the advertising platform violate disparate treatment provisions of the Fair Housing Act?”~\cite{tobin2019hud}. Opening audits up for third party participation could also allow for a variety of stakeholders to come in with diverse yet precise inquiries specific to the context, focus and organizational target of which they are most concerned.

\subsection{Independence} 

The key to third-party audits is independence. Minimizing conflicts of interest and undue influence helps to ensure the reliability of audit results.

\subsubsection{Evidence Base}

\paragraph{\underline{Cross-selling of non-audit services}} Research suggests that auditors who may try to sell additional non-auditing services to their clients are more likely to produce client-favorable audits. Enron’s payment of audit and consulting fees to Arthur Andersen, coupled with Andersen’s fear of losing a very lucrative client, is widely seen to have contributed to the reckless accounting practices that overstated profits by \$1.5 billion~\cite{healy2003fall,mccoy2002realigning}. This 
led to the development of the Public Company Accounting Oversight Board (PCAOB), created by the Sarbanes–Oxley Act of 2002, to oversee financial auditors for public companies.

Causholli et al. (2014) found that financial auditors whose clients express willingness to purchase their non-audit services in the future tend to produce lower-quality audits~\cite{causholli2014future}. Another study found that private car emissions testing facilities with more opportunities to cross-sell services to customers were more likely to pass cars that did not meet emissions standards~\cite{pierce2013role}. Murphy \& Sandino (2009) found that compensation consultants recommended higher CEO salaries when those consultants offered additional services to those executives~\cite{murphy2010executive}. Another study, however, concluded that while cross-selling opportunities could lead to lower quality statutory audits, they did not have an impact on the quality of independent expert reports provided to takeover targets—though the incentive to cross-sell could have an impact on lowering fees~\cite{bedford2021quality}.

\paragraph{\underline{Selection and Compensation}} There is strong social science demonstrating that selection and compensation can undermine independence and, in turn, audit quality. A two-year randomized controlled trial showed that environmental third party audits were more accurate when auditors were not paid by the firm being audited, but by a common pool of government distributed funding~\cite{duflo2013truth}. Similarly, when car owners are given the option to choose their own car inspectors, they select for more lenient car inspectors, resulting in a race to the bottom~\cite{hemenway1990you,bennett2013customer}. One study found that vehicle owners were significantly less likely to keep using car emissions testing stations that fail to report favorably on their vehicles~\cite{bennett2013customer,pierce2013role,hubbard2002consumers}. Similar dynamics have been found in credit rating agencies~\cite{jiang2012does,efing2015structured}, supply chain audits~\cite{short2016monitoring}, and accounting audits~\cite{moore2010conflict}. In short, auditors should ideally not be selected or paid directly by auditees.

\paragraph{\underline{Auditor tenure}} The tenure of auditors is another key determinant of independence~\cite{church2015auditor}. There is strong evidence that auditor-client familiarity can induce auditors to make judgments that are favorable to their client. One analysis of nearly 17,000 supplier audits found that supply chain monitors are more lenient when auditing entities they have inspected before~\cite{short2016monitoring}. Another study found that credit ratings teams were less accurate in their evaluations when they interacted directly with clients~\cite{griffin2011did}. One approach to reduce concerns regarding familiarity is seen in financial auditing, where independence rules prohibit close relationships between auditors and their clients, and prevent the auditor from assisting in the preparation of the financial statements. If an auditor violates these independence rules, regulators can impose a range of consequences ranging from monetary sanctions to a bar from the industry~\cite{sec_leg}.

Some have proposed mandatory audit rotations to counteract this problem. Sarbanes-Oxley, for example, requires mandatory rotation of the lead audit partner. The evidence on mandatory rotations is more mixed. One study observed relatively few benefits to financial audit partner rotations, though it found that the number of financial restatements issued increased in the first two years after rotation, suggesting some “fresh look” benefits~\cite{gipper2021economics}. Other studies focused on the bias-expertise tradeoff~\cite{porter2012complaint,kim2015mandatory,ewelt2012we,houghton2003market,harber2020perceived}. The longer an auditor gets to know a firm, the more firm-specific expertise the auditor acquires. For example, it might take significant time simply to understand the data ecosystem for housing advertisements on a large platform. This tradeoff has also been documented in food safety inspections, where frequent interactions between inspectors and facilities may improve inspector expertise, but can increase the likelihood of social coercion and even bribery~\cite{schuck1972curious,kovacs2020grade}. 

\subsubsection{Implications for AI Auditing}

The AI policy landscape abounds with divergent definitions of “independent audits.” 
The European Commission's Digital Services Act proposal defines “independent'' auditors as being paid and chosen directly by the audit target. 
A New York City Council policy proposal requiring that automated employment screening tools undergo “an impartial evaluation by an independent auditor,’’ fails to even clarify how this auditor is selected and vetted. 

Recent reports have found serious limitations with the voluntary audits undertaken by AI vendors in the hiring space, such as Pymetrics and HireVue. In the case of Pymetrics, 
auditors introduced a framework called the “cooperative audit,’’ which allowed the audit target—Pymetrics—to help frame what questions the auditors would ask, compromising independence. Similarly, Hirevue commissioned and controlled the scope of its audit by ORCAA, including by mediating any public communication of audit results through their own public relations channels. Voluntary audits like these can also be compromised if AI vendors require pre-publication review of audit reports, or if (like HireVue) they mandate that individuals sign an NDA in order to view the audit results. 

Rather than allowing companies to pick their own auditors or dictate the focus of their own audits, the literature suggests that there are significant benefits to a regulator (or third party accreditation body) choosing the auditor or using a central fund, instead of relying on direct payment from audit targets to pay auditors. In particular, the PCAOB experience lends support to the creation of \textbf{\textit{an audit oversight board}}, housed within a public agency or operating as a separate public interest institution, in order to monitor auditors and to determine the expected scope and standards for audit work. In AI, auditors retained by technology companies are potentially compromised by nondisclosure agreements, financial conflicts, and competition for services. An oversight board could define rules around independence, including whether to (a) mandate auditor rotation, (b) discipline  auditors who have violated applicable standards, and (c) prohibit auditors from selling non-audit services to auditees. 

\subsection{Auditor Access}

For third-party audits to be effective, auditors must be granted access to enough data to conduct a robust review. This access can mean physical access to certain premises (as in nursing home inspections) or records and documentation (as in financial audits). 

\subsubsection{Evidence Base}

As Table \ref{ecosystem} shows, third party audits invariably involve privileged access to information. Agencies like the Food and Drug Administration, the National Transportation Safety Board, Federal Trade Commission, and the Consumer Financial Protection Bureau commonly investigate cases of safety risks or performance failures that can harm consumers. Similarly, the Department of Justice and the Department of Housing and Urban Development have engaged in audits to analyze disparate outcomes and compliance to civil rights guidelines for decades~\cite{vecchione2021algorithmic}.

The existing literature has addressed the question of access in a number of ways. Some studies, for example, have examined whether unannounced physical site visits and inspections yield more insight into performance quality than announced inspections, highlighting a concern that audit targets might be able to “game” inspections if they know when they are taking place. One study of 205 NHS hospitals in England found that hospital cleanliness (as perceived by patients) increased during periods when inspections occurred~\cite{toffolutti2017evidence}. Another study found that unannounced inspections provided better information about the quality of care at nursing homes, and could reduce the regulatory burdens of conducting inspections—though announced inspections could be better suited to identifying organizational issues~\cite{klerks2013unannounced}. 

Other research has focused on how differing degrees of audit access influence audit quality. One systematic analysis of the monitoring methods employed by the company PricewaterhouseCoopers (PwC), for example, found that the resulting audits omitted several major violations and contained serious deficiencies~\cite{o2000monitoring}. The author attributed these shortcomings in part to management pre-screening of information and interviews for auditors. Another study found that auditors who were subject to PCAOB inspections produced higher quality audits compared to auditors (typically auditors of foreign SEC registrants) that the PCAOB was not allowed to inspect, suggesting that providing access to parties beyond auditors, such as regulators, may provide additional benefits~\cite{lamoreaux2016does}. Some have also proposed increasing auditor access and visibility into a company’s financial records by updating the AICPA’s Generally Accepted Auditing Standards (GAAS) to allow for data analysis during a financial audit~\cite{titera2013updating}.

\subsubsection{Implications for AI Auditing}

Currently, third parties are often forced to operate at great legal risk to gain access to the audited AI system, keeping them vulnerable to various forms of corporate retaliation or obstruction. For instance, those making use of crowd-sourcing data collection tools can be interrupted abruptly by claims of violations to privacy or intellectual property rights --- NYU researchers studying the Facebook ad delivery algorithm were presented with a cease-and-desist order~\cite{faceshutNYU}; German auditors at Algorithm Watch were legally threatened by Instagram~\cite{faceshutAW}; and civil society watch dogs were subpoenaed by Clearview AI~\cite{shutclearview}. Similarly, several investigators, including academic researcher Christian Sandvig and investigative journalist Julia Angwin, have joined the ACLU to campaign against the use of anti-hacking laws such as the Computer Fraud and Abuse Act (CFAA) to block audit studies that use data scraping~\cite{kadri2020digital, 2020sandvig, 2020van}. Corporations can also use non-legal strategies to block access, including setting up a paywall, adopting prohibitive Terms of Service conditions, or structuring the product to obscure any clear set of test points. To date, it still often requires extraordinary effort for investigators to gain sufficient access for a thorough analysis. The ACLU—representing the National Fair Housing Alliance (NFHA), Communications Workers of America (CWA), several regional fair housing organizations, and individual consumers and job seekers had to demand that Facebook give them access to analyze the discrimination in its customized ad delivery system as part of a legal settlement~\cite{sherwinfacebook}.

Even third parties who are contracted to conduct an audit can run into challenges, as access is provided only on the audited target's terms. For instance, a recent voluntary audit of HireVue, a vendor of algorithmic interviewing and hiring services, was found to be “highly limited”: the audit reviewed HireVue’s documentation of just one of its job candidate assessments, and did not include independent evaluation of HireVue’s data or models. Similarly, attempts by social scientists to work directly with Facebook to gain access to data were thwarted by delayed communications, incomplete data releases and the lack of a clear data dictionary, making the released information indecipherable \cite{fbpersily2018}. As one participant notes, “This was the most frustrating thing I've been involved in, in my life.'' \cite{lapowsky}. Often, such collaborative arrangements also include some control or restriction through contracts of how much of the auditor results can be shared more broadly (as was the case for the Pymetrics audit ~\cite{wilson2021building}).

Lack of access to data and algorithmic systems strikes us as the most significant vulnerability of the current AI audit ecosystem. Protecting proprietary information is not a proper response, as all audit systems provide some form of privileged access to auditors, and disclosure does not need to be direct nor absolute. The National Institute of Standards and Technology, for instance, protects models by having companies run models via a custom Application Program Interface (API) for the Face Recognition Vendor Test (FRVT). Such \textbf{\textit{mediated access}}, subject to auditor vetting (perhaps by an audit oversight board) and consistent with audit scope, will be critical to enabling third party auditing of AI systems. 

\subsection{Professionalization} 
Training, standardization, and accreditation of auditors are also key to ensuring that audits are accurately and consistently conducted. 

\subsubsection{Evidence Base}
Training improves audit quality and more experienced team members are associated with more effective audits~\cite{aobdia2021economics}. One study of supply chain monitors found that audit teams who had more training found significantly more violations than audit teams that were less well-trained~\cite{short2016monitoring}. Another study found that, observing the same conditions, food safety inspectors disagreed on how to implement the health code 60 percent of the time~\cite{ho2017does}. A randomized controlled trial showed that the accuracy and consistency of these inspections improved with training and peer review. Additional research has found that audit quality improves with the amount of client-specific expertise held by audit team members~\cite{aobdia2021economics}.

Other research shows that clear and rigorous standard-setting can improve the quality of third-party audits. Absent standardization, audit variation can lead audited entities to “shop around” for favored auditors~\cite{chung2019opinion}. One study, for example, found that a shift in Belgian financial audit standards towards more rule-based procedures decreased financial auditing errors and increased auditor independence~\cite{carcello2009rules}. Another found that PCAOB inspections, which assess the compliance of registered public accounting firms with professional audit rules and standards, improved audit quality by prompting lower-quality auditors to exit the market~\cite{defond2011effect}. However, studies of accounting standards reveal that tightening the strictness of standards could potentially be a double-edged sword—although effective at aligning auditors with external expectations, overly detailed standards can also lead to complacency by reducing auditor agency and thus promoting a check-boxing “compliance mentality,’’ which disincentivizes skill development and overall individual competence in outlier situations~\cite{gao2019auditing}.  

As seen in the instances where auditees choose auditors in Table \ref{ecosystem}, accreditation is widespread. The FDA, for example, may withdraw an auditor’s accreditation if that auditor certifies a supplier that is later linked to an outbreak of foodborne illness or if it refuses to give the FDA access to its records~\cite{lytton2014oversight}. The literature distinguishes between accreditation, which is typically conducted by a non-governmental body, and licensing administered by the public sector. While much empirical work has focused on the costs and barriers to entry imposed by occupational licensing, less empirical work has probed whether professional licensing has an impact on quality~\cite{kleiner2000occupational}. Existing research on this question is mixed. Several studies have found that increasing or decreasing licensing requirements for CPAs had no meaningful effect on CPA quality~\cite{barrios2021occupational,cascino2021labor,colbert1999state}. Another analysis of dental licensing requirements found that more rigorous licensing standards did not improve average dental outcomes of patients~\cite{kleiner2000does}. One study of occupational licensing laws for teachers found that stricter occupational licensing regimes is associated with a wider distribution of student test scores~\cite{larsen2013occupational}, though others concluded that state occupational licensing requirements for teachers had no impact on students’ standardized test scores~\cite{goldhaber2000does}.    

In general, making audit outcomes accessible to the public through open release or a vetted request process improves auditor accountability. For example, some financial auditors must make their communications and records publicly accessible to shareholders in certain U.S. states—this has been proven to improve auditor conduct and lead to higher quality communications~\cite{turner2010improving}. 

\subsubsection{Implications for AI Auditing}
Given the range of actors who currently provide third party oversight, it will be important to set standards for what counts as an audit and what constitutes acceptable behavior for such a diverse set of potential third party auditors. These auditors make use of a range of tools and methodologies to execute their work, and there is no clear sense of expectations as to what a quality audit entails or who is qualified to execute audit work in the algorithmic context. 

The literature strongly supports \textbf{\textit{training, standardization, and accreditation for third-party AI auditors}}. The literature is much more mixed on the benefits of licensing schemes; accreditation through a professional organization may be preferred for three reasons. First, the need for AI audit services is likely to grow significantly in the near future, and substantial evidence suggests that occupational licensing reduces the supply of services. Second, a fully public inspection system seems unlikely in the near term, as the scale, expertise, and agility needed in the AI governance space far outstrips the current capacity of government agencies \cite{engstrom2020algorithmic}. Third, accreditation is not exclusive of some regulatory oversight. Recall that Enron inspired a re-assessment of Generally Accepted Auditing Standards (GAAS)\footnote{Note that these are standards to judge and guide auditor practice, determined in the US by the Auditing Standards Board of the American Institute of Certified Public Accountants (AICPA). This differs from Generally Accepted Accounting Principles (GAAP), which are standards to guide corporations in their preparation of financial statements.} and prompted greater oversight of auditors by PCAOB. Similar inciting incidents in the medical device industry led to interventions such as the Medical Device Single Audit Program (MDSAP) and to oversight mechanisms for the related Auditing Organizations (AOs)~\cite{chen2018comparative}. 

The capacity gap also strongly calls into question the focus of current policy proposals that focus on academic researchers~\cite{fbpersily2018}.  Certification should extend to a much wider range of potential auditors, including public interest groups, law firms, and journalists. Limiting audit access to academic researchers may narrowly focus audit resources and attention on questions that are important for academic novelty, but misaligned with the needs of other communities. Non-academic groups can also take advantage of a broader range of strategies to conduct audits and understand harms to distinct communities. 

Last, \textbf{\textit{professional standards}} may also determine best practices with respect to legal immunity. Fears of legal repercussions or corporate retaliation can weaken the audit inquiry, and professional standards can help determine limited conditions for liability. Serious violations, such as selling audit target information to competitors, conducting non-audit related activities with the accessed data, or failing to adhere to security or privacy measures, could lead to revocation of credentials. 

\subsection{Post-Audit Actions}
After an audit has been completed, investors, regulators, and other affected individuals may want to access and utilize findings to inform shareholder, enforcement, and private actions. We consider here what steps auditors and auditees should be obligated to take after an audit is completed. We focus particularly on the availability and disclosure of audit results. 

\subsubsection{Evidence Base}

Public visibility of audit results can improve accountability in several ways. It can prevent companies from being able to disguise or hide undesirable audit outcomes~\cite{engler2021independent} and incentivize better behavior. The Insurance Institute for Highway Safety, for example, operates a vehicle safety database that allows users to search for crash test results for various makes and models. The drug company SuperGen saw its stocks plummet after the FDA revealed that it had overstated the efficacy and safety of its cancer drug. One study found that mandated disclosure of food inspection results increased the compliance of food businesses with food safety regulations~\cite{bavorova2017does}, though others have found similar food safety disclosure schemes to be ineffective~\cite{handan2020feasible,ho2012fudging}.  

Separately, the public release of third-party audits can prompt more rigorous auditing practices by allowing for third-party verification. One study found that auditors who have been publicly recognized for their accuracy tend to produce more accurate audits than their non-recognized counterparts, particularly when there are conflicts of interest~\cite{fang2009effectiveness}. An analysis of PwC’s auditing practices also found that the confidential nature of PwC’s audits shielded them from verification by other researchers or NGOs and limited their ability to improve the conditions of the regulated entity~\cite{o2000monitoring}. Securities law is premised on the theory that mandated disclosure will improve corporate behavior, and SEC’s EDGAR database allows users to search corporate filings, find out who audited which company, and access audit reports. Recent changes even provide the name of the audit partner.

Mandatory disclosure requirements, however, are not costless. One study found that publicly traded companies spend an average of \$2.2 million to comply with the Sarbanes-Oxley Act’s financial reporting requirements~\cite{krishnan2008costs}. Another analysis of restaurant grading across ten jurisdictions found that the practice—meant to protect consumers and improve food sanitation via targeted disclosures—suffered from several serious deficiencies, including grade inflation and grading inconsistencies. The implementation of a restaurant grading scheme shifted valuable inspection resources away from more significant safety hazards~\cite{ho2012fudging}. Other commentators have also noted the tension that exists, particularly in the context of algorithmic accountability, between the desire for AI transparency and the desire to enforce existing intellectual property rights. Broadly scoped AI audits—including things like “ethics audits”—could also be expensive to conduct, particularly for newer startups. 

One way of addressing these concerns is to adopt a restricted transparency regime. Sarbanes-Oxley, for example, incentivizes audit firms to fix defects identified in PCAOB inspections by agreeing not to publish deficiencies that are remediated by the firm within 12 months. The FDA reviews applications for medical device premarket approval (PMA), but does not release information about the devices until after a decision has been rendered. Even then, the FDA releases only a summary about the safety and efficacy of the medical device, along with limited data from the PMA file after all sensitive and confidential information has been removed.

\subsubsection{Implications for AI Auditing} 

A restricted transparency regime for registered audit reports could have significant potential benefits. 
To establish public accountability, all audit results should be issued in a report, and such reports registered and made available in summary form (either publicly or upon vetted request) via a database of audit results. A \textbf{\textit{registry of audit results}} avoids the bias from pre-publication review or the secrecy of findings under NDAs and ensures that audits can provide market incentives for companies to correct behavior, inform enforcement actions, and provide an evidence base for standard setting. Such reports would also help cure one of the major emerging informational asymmetries in technology regulation, by opening the black box of algorithmic design. This would also alert regulators directly of audit outcomes, and open communication channels between them and auditors.

Limited disclosure of audit results would balance the protection of proprietary information against the benefits of external accountability. For such reasons, regulators may already be entitled to fuller access and shareholders of publicly traded companies may inspect audit reports under certain circumstances.

\textbf{\textit{Repeated assessments}}, including after product release, can be effective at ensuring compliance. For instance, it was recurring audits that revealed that Facebook was 
continuing to discriminate against minority users in job and housing ads,
despite earlier communications 
that it would no longer do so~\cite{kingsley2020auditing,merrill2020does,ali2019discrimination,kaufman,sapiezynski2019algorithms}.

Audit results can also inform more general \textbf{\textit{standard-setting}} around AI performance and engineering practice. Many AI ethics and accountability discussions have been stymied by the lack of concretely stated expectations of performance and evidence \cite{mittelstadt2019principles}. A major advantage of  the audit approach we articulate is its tangibility: audits can help reveal concrete issues and repeated product failures that can later be resolved by general standards. For instance,  IEEE P7013 Inclusion and Application Standards for Automated Facial Analysis Technology industry standard is directly informed by the Gender Shades audit. 

\section{Limitations}

Here we note several important limitations to our work. First of all, 
our work is only meant to take a first step in the direction of learning from cognate audit schemes, but because the evidence draws from so many different areas, it is necessarily selective and incomplete. 
Additionally, although our article is critical of internal compliance structures, our work is by no means meant to deprecate the value of internal audits, which is also well understood in other industries such as finance~\cite{bedard2010strengthening} 
and instead simply illustrates why we cannot rely on internal audits \emph{alone}.
Next, one might also object that we have commingled public and private inspections with audits, which skews the evidence base.
However, across regulatory schemes, public and private entities functionally audit for the same issues (e.g., product defects); and even the scholarly literature treats inspections as functionally no different than audits (e.g., environmental inspections are described as third party auditors~\cite{duflo2013truth}). 

A final comment is that perhaps we do not go far enough in this article. For instance, while we have synthesized the evidence base around each distinct institutional design element, the actual effect may be interactive. Auditor access may only matter when auditors are also independent. We recognize this and believe future work could fruitfully examine the interaction of these dimensions. Also, due to the fluidity of the space, we have purposely stayed away from suggesting whether these reforms should be promulgated via legislation, regulation, enforcement, or self-regulation, although that is an important next step.
Our point in this specific text is more simple: as the AI community turns toward audits, it has largely ignored what other audit schemes have to teach us on fostering policy environments conducive to third party participation.  

\vspace{-3mm}
\section{Conclusion}
In this article, we demonstrate the dangers of relying on a magic incantation of “audits” to address algorithmic accountability challenges.  
The mere insistence for audits is not enough -- in particular, specific interventions will be necessary to allow for the effective participation of third parties, who play a critical role 
yet continue to face serious and often debilitating challenges when engaged in their work. 
AI policy can foster third party audits with more deliberate institutional design, informed by the audit schemes from other industries. Our specific recommendations point to interventions that can foster an ecosystem in which third parties can not only survive in their role, but thrive in directly confronting, verifying, and subjecting to scrutiny the performance claims made by corporations, while adequately addressing complaints of harm brought forth by the impacted population. 
Our hope is 
to directly contribute to the growth and maturity of a third party audit ecosystem that can provide the thorough external scrutiny required to protect those currently most vulnerable to harm.

\begin{acks}
We thank participants at the 2022 Fall Conference at Stanford's Institute for Human-Centered AI, Fiona Scott Morton, Cathy O'Neil, and DJ Patil for early conversations sparking these ideas, and the Public Interest Technology University Network, the McGovern Foundation, the Algorithmic Justice League, and the Mozilla Foundation for support. 
\end{acks}

\bibliographystyle{ACM-Reference-Format}
\bibliography{sample-base}

\appendix
\section{Appendix}

The following pages include an Appendix, which provides additional context and definitions for the policy survey.

\begin{table*}[!htbp]
\label{ecosystem-features}
\centering
\begin{tabular}{|p{5cm}|p{10cm}|}
\toprule
\textbf{\shortstack{Audit Ecosystem Design Feature}}& \textbf{\shortstack{Explanation}}\\ 
\midrule
Mandatory? & Is the audit required by law or voluntarily undertaken?\\
\midrule
Public or private? & Is the audit conducted by the public sector (e.g. a government agency) or by a private entity?\\
\midrule
Accreditation requirements?& Does the auditing entity have to be accredited, certified, or recognized by an official body before it is allowed to conduct the audit?\\
\midrule
What is being audited?& What is the subject of the audit? Is it a product? Compliance with certain regulations?\\
\midrule
Can the auditor make recs/changes? & Does the auditor have the power to make recommendations or changes to the audited entity?\\
\midrule
Are audits recurring? & Does the audit take place just once? Or is there a periodic auditing requirement?\\
\midrule
Who selects the auditor? & Who selects the entity that will conduct the audit?\\
\midrule
Who pays the auditor? & Who pays the entity that will conduct the audit?\\
\midrule
Public transparency requirements? & Is there a requirement that the contents of the audit be made available to the public?\\
\midrule
Rotation requirements? & Is there a requirement that the auditors rotate off of the auditing process after a certain period of time?\\
\bottomrule
\end{tabular}
\caption{Analyzed Audit Ecosystem Features}
\label{ecosystem-features}
\end{table*}

\begin{table*}
\label{audit-descriptions}
\centering
\begin{tabular}{|p{3cm}|p{14cm}|}
\toprule
\textbf{\shortstack{Auditing program}}& \textbf{\shortstack{Description}}\\ 
\midrule
\midrule
 \multicolumn{2}{c}{\textbf{First Party Audits}}\\
\midrule
EPA Audit Policy  & The Environmental Protection Agency (EPA) issued its "Audit Policy" in 1995. The Policy, which is designed to increase compliance with environmental regulations, eliminates or reduces civil penalties for violations that facilities discover as a result of a self-audit and voluntarily disclose. \\
\midrule
SQA Audits & Software Quality Assurance (SQA) audits are conducted throughout the life cycle of a software program to quality-check the engineering processes and methods used in the software's development. SQA audits are typically conducted in-house, and can include various audit types, including security audits and information systems audits. \\
\midrule
\midrule
 \multicolumn{2}{c}{\textbf{Second Party Audits}}\\
 \midrule
 Supply chain audits  & Because the specific features of second-party audits vary depending on the contractual terms between the auditing and audited entities, this is a case study of a two-year audit of suppliers in the confectionary industry. See \citet{djekic2016improving}.  \\
 \midrule
\midrule
 \multicolumn{2}{c}{\textbf{Third Party Audits}}\\
 \midrule
 SOX financial audits & The Sarbanes-Oxley Act (SOX) requires that companies hire independent auditors to complete yearly checks of their financial reports. A SOX compliance audit also reviews companies' internal controls over financial reporting (e.g. management systems; data security practices) to make sure they are secured against fraud or theft.\\
\midrule
MDSAP & The Medical Device Single Audit Program (MDSAP) allows medical device manufacturers to satisfy the requirements of multiple regulatory jurisdictions with a single regulatory audit of the medical device. The MDSAP certification process takes three years, during which participants undergo annual audits.\\
\midrule
FDA audits of clinical trial data  & The U.S. Food and Drug Administration (FDA) routinely audits clinical trials to confirm the validity of their findings, including trials that may form the basis for FDA drug approval. The FDA also conducts for-cause audits of investigators if someone reports that investigator to the FDA or if the FDA receives a complaint about the investigator from the drug manufacturer. \\
\midrule
California GHG Verification &  According to California state regulations, greenhouse gas (GHG) emissions data reports must be independently checked every year by verification bodies accredited by the California Air Resources Board (CARB).\\
\midrule
USDA National Organic Program & Farms and businesses that want to label a product as "organic" must be certified by an accredited third-party organization. The audit is meant to ensure that the product meets national organic standards. \\
\midrule
CPSC third-party testing program for children’s products & Under federal law, every toy intended for use by children age 12 or under must be tested by an accredited third-party laboratory for compliance with the applicable children's product safety requirements.\\
\midrule
FDA FAERS post-market surveillance &  Under U.S. Federal Communication Commission (FCC) regulations, Telecommunication Certification Bodies (TCB) must audit samples of telecommunications products after they have gone to market (i.e. as part of a post-market surveillance program) to ensure that the products remain in compliance with FCC requirements. \\
\midrule
FCC TCB post-market surveillance & The FDA Adverse Event Reporting System (FAERS) is a computerized database that assists the FDA with its postmarket surveillance activities. Patients, caregivers, and healthcare professionals may submit adverse drug events to FAERS. The reports are assessed by a multidisciplinary staff (and in some cases, by AI/ML tools), after which the FDA may conduct further postmarket studies or take regulatory actions (e.g. updating lables or even reevaluating an approval decision) to improve product safety. \\
\midrule
FAA Office of Audit and Evaluation & The U.S. Federal Aviation Administration (FAA)'s Office of Audit and Evaluation (AAE) receives protected disclosures via its online hotline from FAA employees, certificate holders, an employees of certificate holders concerning potential violations of the FAA's statute and regulations and acts, omissions, or gross misconduct that affect aviation safety. AAE evaluates these disclosures, and if it determines that further investigation is warranted, it will either refer the disclosure to another FAA office or request assistance to conduct an inquiry.  \\
\midrule
HHS nursing home inspections & To participate in the Medicaid and Medicare programs, nursing homes must undergo annual health inspections. These inspections are conducted by state survey agencies, and inspections may take place more often if the nursing home is performing poorly, or if there are complaints or reported incidents within the facility.  \\
\bottomrule
\end{tabular}
\caption{Overview of Select U.S. Audit Programs}
\label{audit-descriptions}
\end{table*}

\end{document}